\documentclass[aps,twocolumn,nofootinbib,amsmath,amssymb,reprint,preprintnumbers,
]{revtex4-1}

\usepackage{dcolumn}
\usepackage[
	  pagebackref=false,
	  colorlinks=true,
      linkcolor=blue,
      urlcolor=blue,
      filecolor=black,
      citecolor=red,
      pdfstartview=FitV,
      pdftitle={},
        pdfauthor={},
        pdfsubject={},
        pdfkeywords={},
        pdfpagemode=None,
        bookmarksopen=true
      ]{hyperref}

\usepackage[normalem]{ulem}
\usepackage{amsmath}
\usepackage{enumerate}
\usepackage{amsfonts}
\usepackage{epsfig}
\usepackage{mathbbol}

\usepackage{color}
\usepackage{ulem}
\usepackage{verbatim}

\newcommand{\be}{\begin{equation}}
\newcommand{\ee}{\end{equation}}
\newcommand{\bea}{\begin{eqnarray}}
\newcommand{\eea}{\end{eqnarray}}
\newcommand{\bega}{\begin{gather}}
\newcommand{\eega}{\end{gather}}
\newcommand{\nn}{\nonumber\\}
\newcommand{\bi}{\begin{itemize}}
\newcommand{\ei}{\end{itemize}}
\newcommand{\ben}{\begin{enumerate}}
\newcommand{\een}{\end{enumerate}}
\newcommand{\bca}{\begin{cases}}
\newcommand{\eca}{\end{cases}}
\newcommand{\bln}{\begin{align}}
\newcommand{\eln}{\end{align}}
\newcommand{\bst}{\begin{split}}
\newcommand{\est}{\end{split}}
\def\ie{\begin{equation}\begin{aligned}}
\def\fe{\end{aligned}\end{equation}}
\newcommand{\bma}{\le(\begin{matrix}}
\newcommand{\ema}{\end{matrix}\ri)}

\newcommand{\gff}{g_{\phi\phi}}
\newcommand{\gtf}{g_{t\phi}}
\newcommand{\rlr}{r_{\text{LR}}}
\newcommand{\tlr}{\theta_{\text{LR}}}

\newcommand{\eqn}{&=&}
\newcommand{\ppa}[2]{\left(\frac{\partial}{\partial #1}\right)^{#2}}
\newcommand{\mg}[1]{\textcolor{red}{\bf [#1---mg]}}
\newcommand{\sg}[1]{\textcolor{blue}{\bf [#1---sg]}}
\newcommand{\eq}[1]{Eq.(\ref{#1})}
\newcommand{\fig}[1]{Fig. \ref{#1}}

\begin{document}

\title{Universal Properties of Light Rings for Stationary Axisymmetric Spacetimes }

\author{Minyong Guo $^{1, 2}$}
\email{minyongguo@pku.edu.cn}

\author{Sijie Gao $^{1}$}
\email{sijie@bnu.edu.cn  (corresponding author)}

\affiliation{ $^1$ Department of Physics, Beijing Normal University, Beijing 100875, China\\
$^2$ Center for High Energy Physics, Peking University, Beijing 100871, P. R. China\\
}
\begin{abstract}
Light rings (LRs) play an important role in gravitational wave observations and black hole photographs. In this paper, we investigate general features of LRs in stationary, axisymmetric,  asymptotically flat spacetimes with or without horizons. For a nonextremal black hole, we show explicitly that there always exist at least two LRs propagating in opposite directions, where the outermost one is radially unstable. For an extremal black hole, we show that there exists at least one retrograde LR.  We find that there is at least one LR which is angularly stable. The stability analysis does not involve any energy condition. Our method also applies to horizonless spacetimes and we prove that LRs always appear in pairs.  Only some natural and generic assumptions are used in our proof. The results are applicable to general relativity as well as most modified theories of gravity. In contrast to previous works on this issue, we obtain much stronger results with a much more straightforward approach.
\end{abstract}

\maketitle

\section{Introduction}
In recent years, many evidences, including gravitational wave signals from the merger of binary system observed by LIGO and Virgo \cite{Abbott:2016blz, LIGOScientific:2018mvr} and the image of M87* photographed by the Event Horizon Telescope (EHT) \cite{Akiyama:2019cqa, Akiyama:2019brx, Akiyama:2019sww, Akiyama:2019bqs, Akiyama:2019fyp, Akiyama:2019eap}, strongly support the existence of astrophysical black holes in our universe. Therein, the location of the last radially unstable circular photon orbit  is necessary to describe the dynamics of the binary system from the inspiral phase to the ring-down \cite{Buonanno:2000ef, Cardoso:2016rao, Cunha:2018acu, Barack:2018yly}. More precisely, in the ringdown process, the black hole dissipates through a set of quasinormal modes leaving the gravitational wave footprint of the event horizon, and the light ring (LR) is the outer boundary for the quasinormal modes \cite{Maggio:2020jml, Cook:2020otn}. One expects gravitational wave ringdown signals would give a generic feature of any dynamical scenario, the existence of the light ring is indispensable as the theoretical basis.

The size of a stationary black hole is closely connected with its photon region, which is the set of all radially unstable bound photon orbits and can be seen as a border between light rays that escape to the far region and light rays that fall into the black hole \cite{synge, bardeen, Compere:2020eat, Compere:2019cqe, Johnson:2019ljv, Wang:2020emr, Gralla:2019xty, Peng:2020wun, vitor14, Cunha:2017eoe}. For example, the well-known photon orbits with the radius $r=3M$ in the Schwarzschild spacetime constitute the border between trapped and untrapped light rays. Among them, circular photon orbits are called light rings. It is worth mentioning that  observable light rings must be unstable in the radial direction, since these photons would fall into the black holes or escape to the infinity once perturbated. On the contrary, if LRs are stable in the radial direction, they would not be captured by telescopes and the images of black holes would not be obtained. As a result, the radially unstable LRs play an important role in both gravitational wave detection and black hole shadow observation. Therefore, the study of radially unstable LRs is essential for black hole theories and observations.

Although LR has been extensively discussed in recent years, most of the works focus on specific spacetimes.  A major development in this research area was made in \cite{prl17}. By using an elegant topological argument, the authors found LRs outside a horizonless ultracompact object (UCO) always come in pairs. Very recently, the authors employed this topological argument to a stationary, axisymmetric, asymptotically flat black hole spacetime, and  found that at least one standard LR exists outside the nonextremal horizon for each rotation sense \cite{prl20}. Along this line, an extension to spherically symmetric black hole with AdS and dS behaviors has been made in \cite{Wei:2020rbh}. These findings have greatly improved our understanding on LRs. However, there are still important issues that remain unclear and unresolved. Firstly, in \cite{prl17}, the topological argument is only suitable for those UCOs which form dynamically from gravitational collapse, starting from an approximately flat spacetime. However, there is no general proof that a UCO must form in this way, although it has been shown that some boson stars can  be formed dynamically from a spherically symmetric process of gravitational collapse and cooling \cite{prl94}, as well as Proca stars \cite{plb16, DiGiovanni:2018bvo}.
  In addition, the topological argument is subtle because it requires a deformation of a sequence of off-shell spacetimes that possess the two Killing vector fields \cite{Cunha:2018acu}. It is not clear whether this can be done in general. Thus, it's better to give a direct proof which requires no knowledge of the history of UCO formation. Secondly, a nice formula was derived in \cite{prl17} which relates the null energy condition to the stability of the LR. However, this formula is a combination of the radial and angular directions. It cannot answer whether the radial or the angular direction is stable.  In particular, the radial stability of LRs is very important  for  the  black hole shadow or the ring-down phase of a binary system. Thirdly, the topological argument cannot predict the existence of LRs on the  equatorial plane when the spacetime possesses the parity reflection symmetry.
Finally, the argument in \cite{prl20} depends on the assumption that the black hole is nonextremal, which does not apply to an extremal black hole.

In this paper, we address all the unsolved issues above and obtain satisfactory answers. In \cite{prl20}, the existence of a standard LR was proved, which means that a LR is a saddle point in the  $r-\theta$ plane. To make a more precise prediction, we
introduce the normal light ring (NLR) to describe a light ring which is radially unstable and angularly stable. By analyzing the generic behaviors of a stationary, axisymmetric, asymptotically flat black hole, we show that there always exist at least two NLRs propagating in opposite directions outside the nonextremal horizon. For extremal rotating black holes, we  find there exists at least one retrograde NLR.  By applying our method to horizonless spacetimes, which represent untracompact objects, we recover the previous result that LRs always come in pairs \cite{prl17}. Obviously, the existence of NLR is a stronger result than the previous standard LR. More importantly, our proof involves only the stationary solution, independent of its history of formation. Our argument can also guarantee the existence of LR on the equatorial plane when the spacetime possesses a reflection symmetry.

The paper is organized as follows. In section \ref{section2}, we introduce some properties  of null geodesics in axis-symmetric spacetimes. In section \ref{section3}, we prove the existence of LRs in axis-symmetric black holes. In section \ref{section4}, we study LRs in axis-symmetric horizonless spacetimes. Conclusions and discussions are given in section \ref{section5}.

\medskip

 \section {Axis-symmetric spacetime and null geodesics.}\label{section2}
 Let us start with a stationary spacetime  described by the metric $ds^2=g_{tt}(r, \theta)dt^2+g_{rr}(r, \theta)dr^2+2g_{t\phi}(r, \theta)dt d\phi+g_{\theta\theta}(r, \theta) d\theta^2+g_{\phi\phi}(r, \theta) d\phi^2 $, where we have employed the circularity of the spacetime, which gives $g_{r\theta}=0$ under  a suitable gauge \cite{noncircular}. We also assume that the metric is at least $C^2$-smooth \cite{prl17}. In terms of the coordinate system $\{t, r, \theta, \phi\}$, the Killing vectors can be represented by $\partial_t$ and $\partial_\phi$. In general, the 4-momentum of a photon is written as
\be
p^a=\dot t \ppa{t}{a}+\dot r\ppa{r}{a}+\dot{\theta}\ppa{\theta}{a}+\dot\phi\ppa{\phi}{a} \,,
\ee
 where the dot denotes the derivative with respect to an affine parameter. The Killing vectors $\partial_t$ and $\partial_\phi$ give us the conserved energy and angular momentum
\bea\label{EE}
E\eqn -g_{ab}p^a\ppa{t}{a}=-g_{tt}\dot t-g_{t\phi}\dot\phi \,,\\
L \eqn g_{ab}p^a\ppa{\phi}{a} =g_{\phi\phi}\dot\phi+g_{t\phi}\dot t  \,,\label{LL}
\eea
respectively. Since $p^ap_a=0$ along a photon's trajectory, we have
\be
g_{tt}\dot t^2+g_{rr}\dot r^2+2\gtf\dot t\dot\phi+g_{\theta\theta}\dot{\theta}^2+\gff\dot\phi^2=0\,.
\ee
Combining Eqs.(\ref{EE}) and (\ref{LL}), it is not difficult to find
\be
g_{rr}\dot{r}^2+g_{\theta\theta}\dot{\theta}^2+V(r, \theta)=0\,,
\ee
where $V(r, \theta)=-\frac{1}{D}\left(E^2\gff+2EL\gtf+L^2g_{tt}\right)$ is the effective potential with $D\equiv \gtf^2-g_{tt}\gff$. For convenience, we introduce a new parameter

\be
\sigma=\frac{E}{L}\,,
\ee
which is the inverse of the familiar impact parameter. Thus the effective potential can be rewritten as \cite{prl17}
\be\label{potential}
V=-\frac{L^2\gff}{D}\left(\sigma-H_+\right)\left(\sigma-H_-\right)\,,
\ee
where
\be\label{potentialH}
H_{\pm}=\frac{-\gtf\pm\sqrt{D}}{\gff} \,.
\ee
Then, we have
\be
H_+-H_-=\frac{2\sqrt{D}}{g_{\phi\phi}}
\ee
is always non-negative. In addition, if spacetimes of interest contain black holes, denoting the horizon radius as $r_h$, one can show \cite{prl20, Medved:2004tp}
\bea\label{horizon}
D|_{r_h}=0\,,
\eea
and $D$ is always positive outside the horizon since $D$ is the determinant of the  $t-\phi$ sector of the metric \cite{prl20}. Thus $H_+=H_-$ only occurs on the horzions of black holes and $H_+>H_-$ is always true for UCOs.

Combining  Eqs. (\ref{potential}) and (\ref{potentialH}), we find the LRs occur at \cite{Cunha:2016bjh}
\be\label{conHp}
\partial_m H_+=0 \quad \text{and}\quad \sigma=H_+(\rlr, \tlr)\,,
\ee
or
\be\label{conHm}
\partial_m H_-=0 \quad \text{and}\quad \sigma=H_-(\rlr, \tlr)\,,
\ee
where, $m\in\{r, \theta\}$ and $(\rlr, \tlr)$ represent the  coordinates of the LR.

Note that for the LR associated with $H_+$, we have
\bea
\partial_m^2V(\rlr, \tlr)=\frac{L^2\gff}{D}\partial_m^2H_+(\rlr, \tlr)(H_+-H_-) \,\nn.
\eea
For the LR associated with $H_-$, we have
\bea
\partial_m^2V(\rlr, \tlr)=\frac{L^2\gff}{D}\partial_m^2H_-(\rlr, \tlr)(H_--H_+) \,\nn.
\eea
Since $H_+>H_-$ for any r that is not on the horizon, we see that  $\partial_m^2H_+$ has the same sign as $\partial_m^2V$ and $\partial_m^2H_-$ has the opposite sign. This property will be used in the stability analysis.

Now we are ready to explore the nature of LRs in axis-symmetric black holes and horizonless spacetimes.

\medskip

\section{Axis-symmetric black holes}\label{section3}
\begin{figure}[htbp]
\centering
\includegraphics[width=8.5cm]{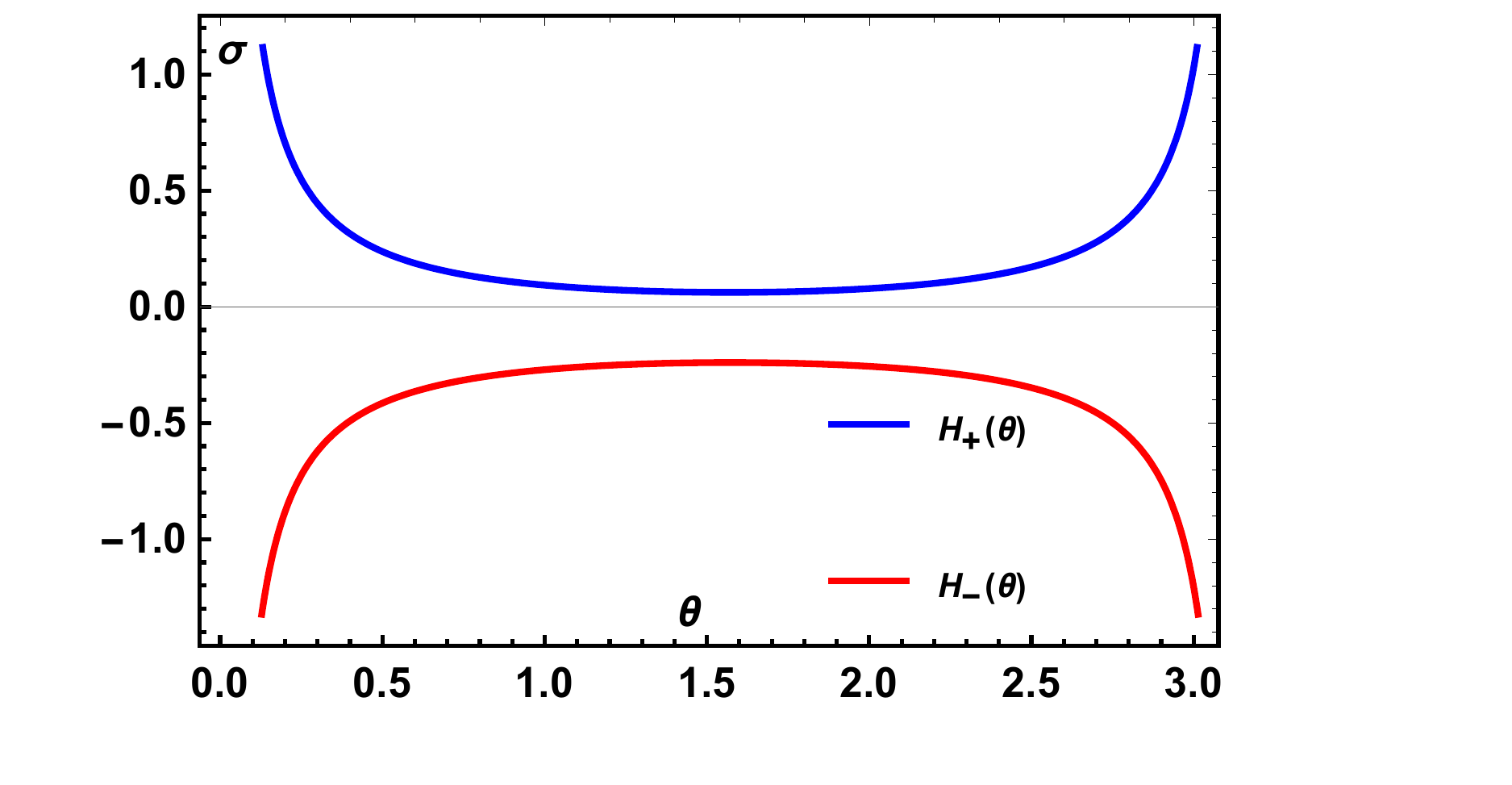}
\caption{The functions $H_+(r, \theta)$ and $H_-(r, \theta)$  for Kerr black hole.  We choose $a=-0.5$ to ensure $\gtf(r_h)>0$. The minimum of $H_+$ and the maximum of $H_-$ both correspond to minimum values of the potential $V$ in the $\theta$ direction. }
\label{diagt}
\end{figure}

We first consider axis-symmetric black holes. Let us start with the angular direction, i.e., $m=\theta$. Near the axis, as shown in \cite{prl20}, $\rho\equiv\sqrt{g_{\phi\phi}}$ goes to zero when $\theta\to0$ and $\theta\to\pi$. Also, we have
\begin{align}
\begin{cases}
\partial_\theta\rho>0 & \text{  $\theta\to0$}\,,\\
\partial_\theta\rho<0 & \text{  $\theta\to\pi$}\,.\\
\end{cases}
\end{align}
Thus we find
\bea
H_\pm\simeq\pm\frac{1}{\rho}\to\pm\infty\,,
\eea
and
\begin{align}
\partial_\theta H_\pm\sim\mp\frac{\partial_\theta\rho}{\rho^2}\sim
\begin{cases}
\mp\infty & \text{  $\theta\to0$}\\
\pm\infty & \text{  $\theta\to\pi$},\\
\end{cases}.
\end{align}
So, for any fixed $r$, $H_\pm$ can be viewed as a function of $\theta$ ranging from $0$ to $\pi$, as shown in \fig{diagt}.
This means that for each given  $r>r_h$,  there always exists a $\theta=\theta_+$  such that $H_+(r,\theta_+) $ is a minimum in the $\theta$ direction. In this way, we obtain a function $\theta=\theta_+(r)$.  Similarly, we have $ \theta_-(r)$ for $H_-$.

 In asymptotically flat spacetimes,  $H_\pm\to \pm\frac{1}{r\sin\theta}$ as $r\to\infty$.
Thus, we have
\be
\theta_\pm(r\to\infty)=\pi/2 \,. \label{sinf}
\ee

Next, we turn to the radial direction. For any fiexd $\theta$, $H_\pm$ can be viewed as functions of $r$.
To investigate general features of LRs, it is crucial to analyze the behaviors of  $H_\pm$ at infinity and the horizons. For asymptotically flat spacetimes, when $r\to\infty$, we have $\gtf\to 0$, $g_{tt}\to-1$ and $\gff\to r^2$. Thus, $H_\pm \to 0^\pm$.

\begin{figure}[htbp]
\centering
\includegraphics[width=8.5cm]{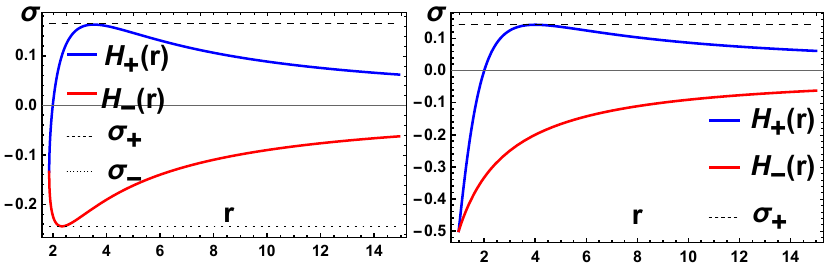}
\caption{The functions $H_+(r)$ and $H_-(r)$  for Kerr black holes. On the left panel, we choose $a=-0.5$ to ensure $\gtf(r_h)>0$, and set $M=1$ so that the Kerr black hole is nonextremal.  On the right panel, we choose $M=1$ and $a=-1$, which corresponds to an extremal Kerr black hole. By imposing the boundary conditions, it is easy to see that there are two radially unstable LRs for the nonextremal Kerr black hole. However, there is only one LR, associated with $H_+$, for the extremal black hole. }
\label{diag}
\end{figure}

Let us turn to the horizons and without loss of generality we can set
\be
g_{t\phi}(r_h)>0\,.
\ee
Then, from Eqs. (\ref{potentialH}) and (\ref{horizon}), we have
\bea
H_\pm|_{r_h}=-\frac{\gtf}{\gff}\Big|_{r_h}<0\,. \label{hpp}
\eea
Note that $H_+\rightarrow 0^+$ as $r\rightarrow\infty$ if $\theta\neq 0 $ or $\pi$. We see immediately that $H_+$ must change sign when approaching the infinity and possess at least one maximum outside the horizon (see \fig{diag}). Differing from the argument in \cite{prl20}, our result holds for both nonextremal and extremal black holes.

The argument for $H_-$ is not so simple because $H_-$ is negative at $r=r_h$ and $r\to\infty$. Note that
\bea
&&\partial_rH_-(r_h)\sim-\frac{\partial_rD(r_h)}{2\sqrt{D(r_h)}\gff(r_h)}\nn
&+&\frac{\gtf(r_h)\partial_r\gff(r_h)-\partial_r\gtf(r_h)\gff(r_h)}{\gff^2(r_h)}\,,
\eea
where we have used $D(r_h)=0$. One can show that $\partial_rD(r_h)>0$ for nonextremal horizons and $\partial_rD(r_h)=0$ for extremal horizons \cite{prl20,Medved:2004tp}. Thus,
\be
\partial_rH_-(r_h)\to-\infty
\ee
for non-extreme black holes, which means that $H_-$ must possess a minimum outside the horizon for any constant $\theta$. However, the sign of $\partial_rH_-(r_h)$ is undetermined for extremal black holes and thus the minimum for $H_-$ may not exist in general. The above results  are clearly illustrated in Fig. \ref{diag}, which depicts $H_\pm$ for extremal and nonextremal Kerr spacetimes.

\begin{figure}[htbp]
\centering
\includegraphics[width=8.5cm]{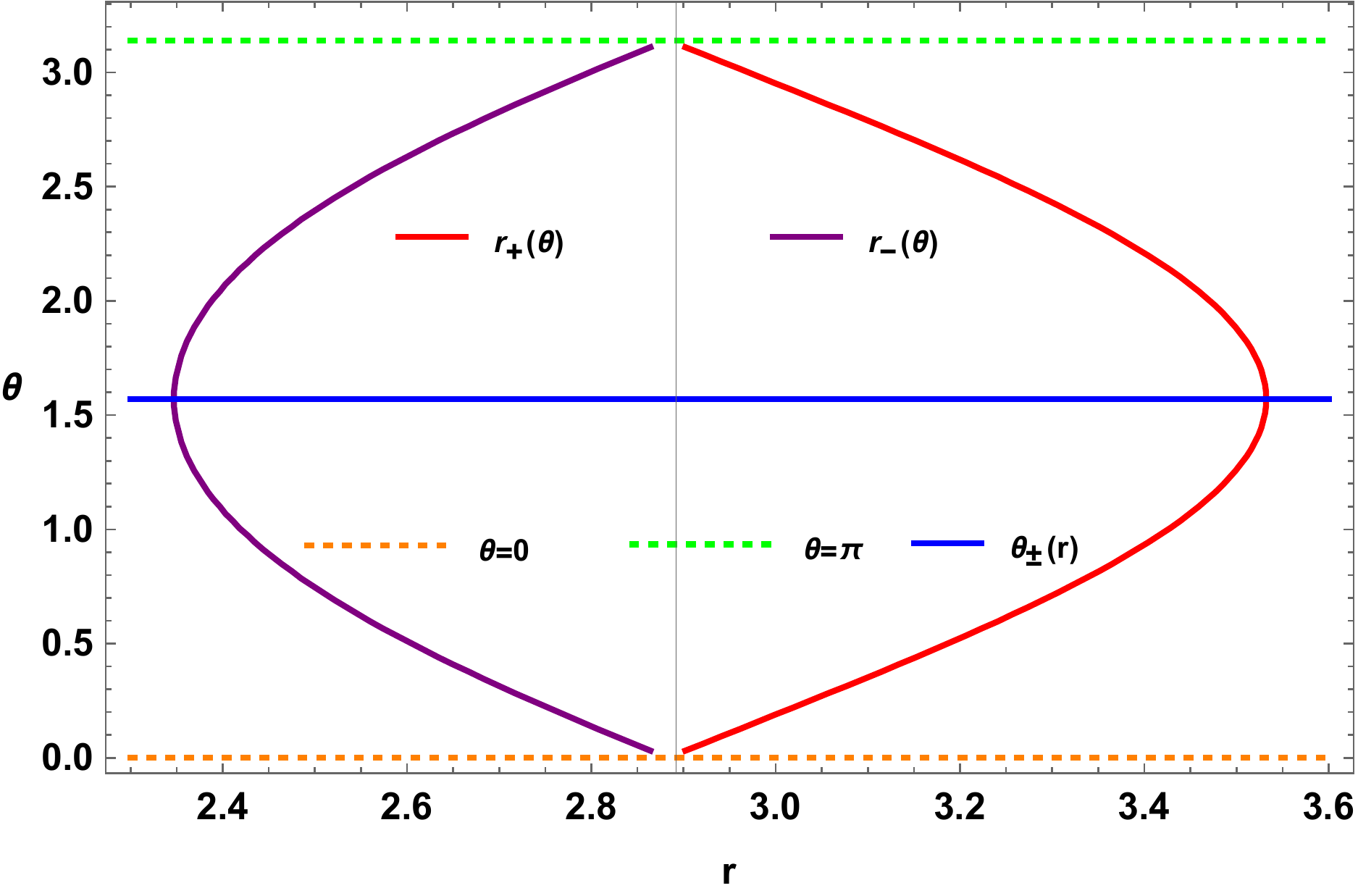}
\caption{The functions $r_\pm(\theta)$ and $\theta_\pm(r)$  for the Kerr black hole with $a=-0.5$. In particular, due to the reflection symmetry, $\theta_+(r)=\theta_-(r)=\pi/2$ is a constant. In general,  $\theta_\pm$ is not necessarily a constant. The intersection points between $r_\pm(\theta)$ and $\theta_\pm(r)$ are NLRs as expected.}
\label{cross}
\end{figure}
\medskip

The above argument holds for any constant $\theta$ in the range $(0,\pi)$. So for each $\theta$, there exists  $r=r_+>r_h$ such that $H_+$ takes an maximum value in the $r$ direction. Hence, we have a function  $r_+(\theta)$ defined in $0<\theta<\pi$.

As we have discussed, a LR exists for the $H_+$ branch if $\theta_+(r)$ and $r_+(\theta)$ intersects on the $r-\theta$ plane. It is reasonable to assume that $\theta_+(r)$ and $r_+(\theta)$ are continuous functions since spacetimes of interest possess $C^2$-smooth. Note that roughly $r_+(\theta)$ divides the $r-\theta$ plane into two regions \footnote{These two regions might be connected at the boundary $\theta=0, \pi$ because $r_+(\theta)$ may not exist there.}. We label the two regions by $I$ and $II$, where region $I$ is on the ``left'', containing the coordinate origin. Since the function $\theta_+(r)$ is defined on $r_h\leq r\leq \infty$, $\theta_+(r_h)$ must lie in region $I$ and from $\theta_+(\infty)=\pi/2$ (see \eq{sinf}), we can conclude that $\theta_+(\infty)$ must lie in region $II$. By  continuity, we see immediately that the two curves must intersect at some point, which is just the location of the LR. One may think that the intersection can be avoided if $\theta_+(r)$ passes through $\theta=0$ or $\theta=\pi$. However, this cannot happen because the minimum value of $\theta$  cannot be $0$ or $\pi$  as demonstrated in \fig{diagt}. The functions $r_+(\theta)$ and $\theta_+(r)$ are plotted  in Fig. \ref{cross}, taking the Kerr black hole as an example. Note that the LR is unstable in the radial direction and stable in the angular direction, we conclude that the LR is a NLR. In addition,  since $H_+(r)$ becomes positive away from an ergosphere, we see from  \fig{diag} that there must be a NLR  lying outside the ergosphere.  The above results hold for $H_-$ if the black hole is nonextremal.

When the spacetime adimits the reflection parity symmetry, it is easy to see that $\partial_\theta H=0$  on the equatorial plane $\theta=\pi/2$. Thus, $\theta_\pm(r)=\pi/2$, which means that there always exist LRs on the equatorial plane.  \\

\begin{figure}[htbp]
\centering
\includegraphics[width=8.5cm]{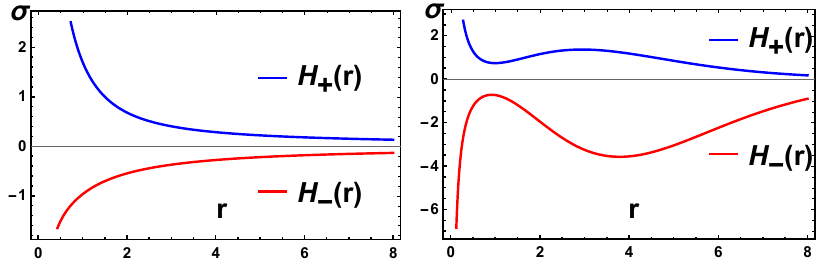}
\caption{The illustrated functions $H_+$ and $H_-$ with respect to $r$ for axis-symmetric horizonless spacetimes. From the asymptotic behaviors at $r=0$ and infinity, we see that there are either no LRs, as shown on the left panel, or at least two LRs, as shown on the right panel.  The inner LR is stable in the radial direction and the outer one is unstable.}
\label{uco}
\end{figure}

\begin{figure}[htbp]
\centering
\includegraphics[width=6.5cm]{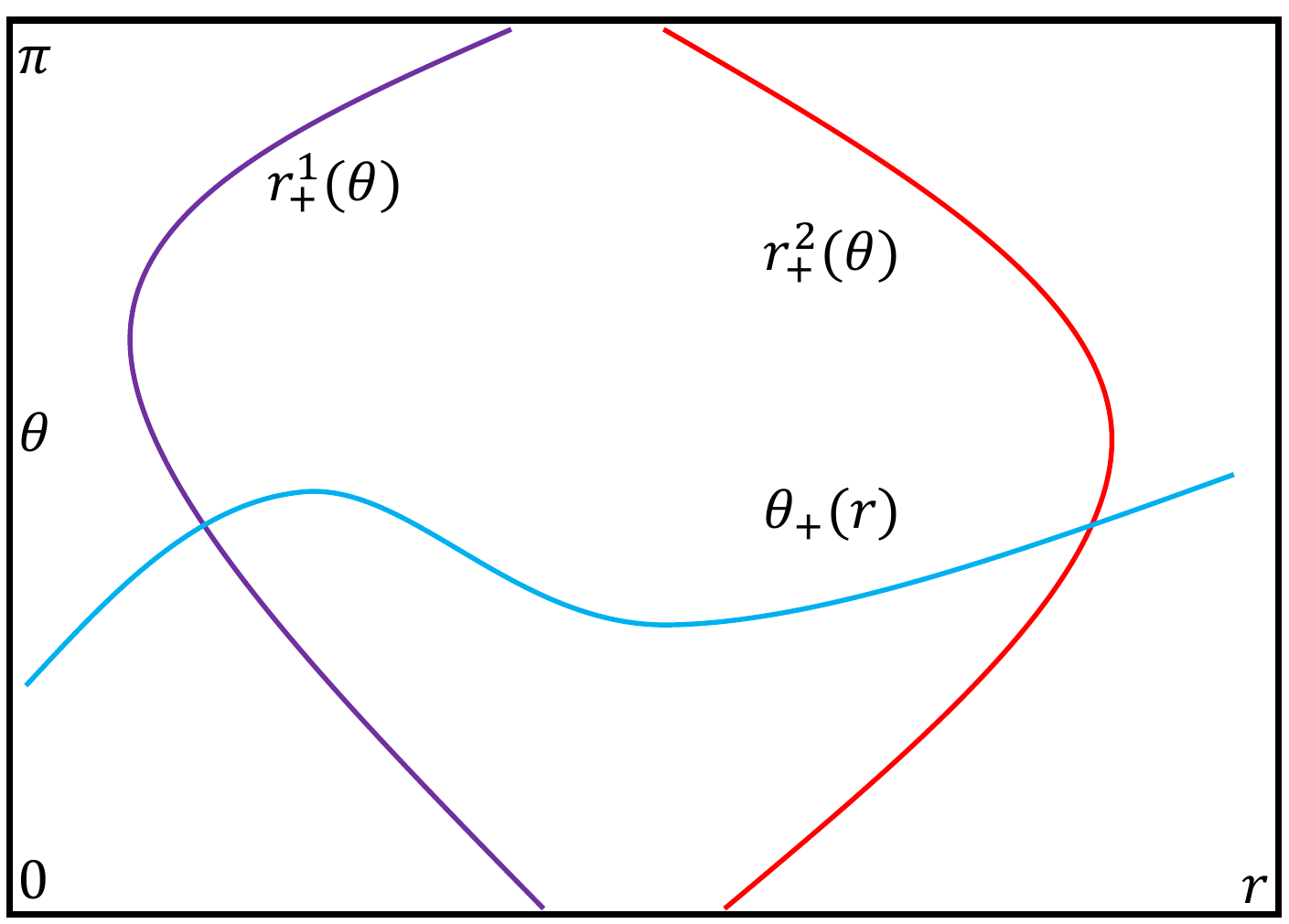}
\caption{An illustration of $r_+(\theta)$ and $\theta_+(r)$ for UCOs.   $r^1_+(\theta)$ and $r^2_+(\theta)$ are local minimums and maximums of $H_+(r)$, respectively. The intersections correspond to LRs.}
\label{crossuco}
\end{figure}

\section {Axis-symmetric horizonless spacetime}\label{section4}
Horizonless spacetimes are interesting because they can represent ultracompact objects (UCO). First of all, it is easy to find that the behaviour of $H_\pm(r, \theta)$ for a UCO is the same as those of a black hole along the angular direction, i.e., there always exists minimum or maximum in the $\theta$ direction for $H_+$ or $H_-$.
 As for the radial direction, the behavior of $H_\pm(r)$ at infinity obviously is the same as for black holes, i.e., $H_\pm\to 0^\pm$. Since the horizon is absent, we can assume that the metric components are regular everywhere. In addition, we assume $\gtf>0$ and $g_{tt}<0$ for $r\to0$. This assumption excludes the case that an ergosphere exists at the center of a UCO, but still can cover a wide class of solutions.

Now we need to focus on the behavior of $H_\pm$ near the  center of the equatorial plane. Let $r$ be the areal radius such that $r\to 0$ and $\gff\to r^2$ near the center. For a UCO,  the metric should be regular at the center (see an example in \cite{Delgado:2020udb}).  Thus we can assume that $g_{tt}\to -k^2$, where $k\neq 0$ is a constant, and $\gtf\to pr^s$ where $p$ is a positive constant and $s\ge0$. Then, near the center we have
\bea
H_\pm\sim \frac{-pr^s\pm\sqrt{p^2r^{2s}+k^2r^2}}{r^2}\,.
\eea
For $s>1$, we can drop the $r^{2s}$ term in the square root and find
\bea
H_\pm\sim \frac{-pr^s\pm\sqrt{k^2r^2}}{r^2}\sim \pm\frac{1}{r}\to\pm\infty\,.
\eea

For $s=1$, it's not hard to see
\bea
H_\pm\sim\frac{\pm\sqrt{p^2+k^2}-p}{r}\to \pm\infty \,.
\eea

For $0<s< 1$, we find
\bea\label{Hpb}
H_+\sim\frac{-pr^s+pr^s\left(1+\alpha^2r^{2-2s}\right)^{1/2}}{r^2}\sim \frac{\alpha^2p}{2}r^{-s}\to \infty\nn
\eea
for a positive $s$, where we have defined $\alpha\equiv\frac{k}{p}$. However, for the branch $H_-$, we have
\be\label{Hmb}
H_-\sim -\left(\frac{2}{r^{2-s}}+\frac{\alpha^2}{2r^s}\right)p\to -\infty \,.
\ee
Thus, for $s>0$, we always have $H_\pm\to \pm\infty$ as $r\to 0$ and $H_\pm\to 0^\pm$ as $r\to\infty$. This implies, from the smoothness of $H_\pm$, that the maximum and minimum of $H_\pm$ must appear in succession.

Finally, for $s=0$, we have
\be
H_\pm\sim\frac{-p\pm\sqrt{p^2+k^2r^2}}{r^2}=\frac{-1\pm\sqrt{1+\alpha^2r^2}}{r^2}p\,,
\ee
as $r\to 0$, which means $H_+\to\frac{\alpha^2}{2}p>0$ and $H_-\to-\infty$.
So we need further analyze the behavior of $H_+$ near $r=0$. It is easy to see
\be
\partial_rH_+(0^+)=0,\quad\text{and}\quad \partial_r^2H_+(0^+)=-\frac{\alpha^4}{4}p<0,
\ee
Hence, $\partial_rH_+$ become negative just away from $r=0$, which, again, indicates that possible extreme values of $H_\pm$ must appear in pairs. The above results are illustrated in \fig{uco}.


Combining the radial and angular directions with the arguments similar to the black hole case,
 we can conclude that for a stationary, axis-symmetric horizonless spacetime with regular metric functions, $H_+$ or $H_-$ always leads to even number of LRs, where the inner one is stable in the radial direction and the outer one is unstable. So the outer one is a NLR. An illustration is given in Fig. \ref{crossuco}.

\medskip

\section{Discussion}\label{section5}
To summarize, we have shown that there are at least two NLRs outside a nonextremal stationary black hole, corotating or counterrotating with the horizon.  We demonstrated that the counterrotating NLR must lie outside the ergosphere. For an extremal stationary black hole, we find there is at least one counterrotating NLR. For horizonless spacetimes, we have proved that if LR exists, there are at least two NLRs, with the outer one being unstable in the radial direction and the inner one being stable. In our arguments, only some generic conditions have been used, for instance, the asymptotically flat condition and the behaviors of the metric near the horizon or the center of a star. These results could play an important role in gravitational wave observations and shadow imaging of the Event Horizon Telescope.

Compared to the theorems proposed in \cite{prl17, prl20}, we have made significant improvements in the following aspects.  First of all, our proof  requires no knowledge of the history of UCO formation, unlike the argument in \cite{prl17}. Moreover, our proof can guarantee the existence of LR on the equatorial plane if there exists one. The Cunha-Herdeiro theorem in \cite{prl20} predicts that one LR outside the black hole must be a saddle point on the $(r,\theta)$ plane. But the stability for the $r$ or $\theta$ direction alone is unclear. In contrast, we have analyzed the stability in each direction. We found that the outermost LR is always unstable in the radial direction and stable in the angular direction. Our results of stability are more specific and do not rely on any energy condition. We also considered the presence of ergosphere and proved that the retrograde LRs must appear outside the ergosphere. Finally, our argument, after some modifications, can naturally apply to extremal horizons.  However, for the extremal case, we can only guarantee the existence of one singe LR which means that the photon shell, which is bounded by the two LRs in the nonextremal case, may not exist. Consequently, the closed shadow curves might not exist at all or exist only for observers with some inclination angles. It is possible that the second LR can be found for some extremal black holes and we shall leave this issue to future works.



\medskip

\section*{Acknowledgments}
SG is supported by NSFC Grants No. 11775022 and 11873044. MG is supported by NSFC Grant No. 11947210. MG is also funded by China Postdoctoral Science Foundation Grant No. 2019M660278 and 2020T130020.


\end{document}